\documentclass[preprintnumbers,amsmath,amssymb,nofootinbib
,superscriptaddress]{revtex4}

\usepackage{hyperref}
\usepackage{graphicx}
\usepackage{color}
\usepackage{xcolor}
\usepackage{comment}
\usepackage{lmodern}
\usepackage{amssymb}
\usepackage{lipsum}
\usepackage{mwe}

\newcommand{\backo}{\!\!\!\!\!\!\!\!\!\!}
\newcommand{\ebo}{\eea }
\newcommand{\bbo}{\bea  && }

\newcommand{\bl}{\biggl(}
\newcommand{\br}{\biggr)}

\newcommand{\vvr}{\vec{r}}

\newcommand{\vvq}{\vec{q}}

\newcommand{\be}{\begin{equation}}  
\newcommand{\ee}{\end{equation}}  
\newcommand{\bea}{\begin{eqnarray}}   
\newcommand{\eea}{\end{eqnarray}}  
\newcommand{\ba}{\begin{array}}  
\newcommand{\ea}{\end{array}}

\usepackage{diagbox}

\newskip\humongous \humongous=0pt plus 1000pt minus 1000pt

\newif\ifdtup

\def\oldreffmt#1{\rlap{[#1]} \hbox to 2\parindent{}}

\def\figfmt#1{\rlap{Figure {#1}} \hbox to 1in{}}  
  
\def\slash#1{#1\!\!\!\!/\!\,\,} 	
\def\beq{\begin{equation}}  
\def\eeq{\end{equation}}  
\def\bea{\begin{eqnarray}}  
\def\eea{\end{eqnarray}}  
\def\half{\frac{1}{2}}  
  
\def\bq{\begin{quote}}  
\def\eq{\end{quote}}

\def\half{\frac{1}{2}}    
\newcommand{\nonumbo}{ \nonumber \\ && }

\relax

\newdimen\tdim  
\tdim=\unitlength  
\def\bar{\overline}

\RequirePackage{caption}
\begin{document}

\preprint{Delafield-1220-2025}
\title{
The Nontrivial Vacuum Structure\\
of an Extended $t\bar{t}$ BEH (Higgs) Bound state \\
}
\author{Christopher T. Hill}
\email{hill.delafield.physics@gmail.com}
\affiliation{Fermi National Accelerator Laboratory,
P. O. Box 500, Batavia, IL 60510, USA}
\affiliation{Department of Physics, University of Wisconsin-Madison, Madison, WI, 53706
}

\begin{abstract}
 
In a recent reformulation of top-quark condensation for the Brout-Englert-Higgs boson, 
we introduced an extended internal wave-function, $\phi(r)$. We show how
this leads to  a {\em manifestly} Lorentz invariant formalism,
where the absence of ``relative time'' is a gauge invariance of the bilocal field theory.
This dictates a novel and nontrivial Lorentz invariant vacuum structure for the BEH boson, 
the relativistic generalization of a condensed matter state analogous to a BCS condensate. 

\end{abstract}

\maketitle
 
\date{\today}


\email{  }

\vspace{-0.25in}
\section{Introduction }

Beginning with Schr\"odinger, any two-body bound state can be  described in a semiclassical limit as a bilocal field, $H(x,y)$   \cite{Schrodinger,Yukawa}.
For the ground state ansatz this factorizes in barycentric (center-of-mass system) coordinates as:
\bbo
H(x^\mu,y^\mu) \sim H(X^\mu)\phi(r^\mu).
\ebo
In the case of a Brout-Englert-Higgs (BEH) boson, composed of $\bar{t}t$, massless chiral top and anti-top quarks 
are located
at space-time coordinates $(x^\mu,y^\mu) = X^\mu\pm r^\mu$. 
$H(X^\mu)$ can be viewed as the standard model (SM)  BEH  isodoublet with
electroweak charges and $\phi(r^\mu)$ is a complex scalar which is electroweak
neutral \cite{chris1,chris2}.  $H(X^\mu)$ describes the center-of mass motion of
the BEH boson, and $\phi(r^\mu)$ is then the internal wave-function.


A theory of a BEH boson composed of $\bar{t}t$, known as ``top condensation,'' 
was proposed in the 1990's \cite{Yama,BHL,Topcolor,NSD},
deploying the Nambu-Jona-Lasinio (NJL) model \cite{NJL}\cite{BHL}. 
The NJL model, however, is pointlike, lacking $\phi(r^\mu)$, which leads to difficulties
when there is a large hierarchy between the composite scale, $M_0$, and the electroweak scale $|\mu|$ 
(the symmetric phase mass of the BEH boson, $|\mu|=88$ GeV). 
The inclusion of the internal wave-function 
yields significant improvement in the predictions of
the low energy parameters.
In the  limit $|\mu|<\!\!< M_0$ there is
significant wave-function spreading of $\phi(\vec{r})$ and the resulting dilution effects dominate
the low energy effective theory.  This brings its predictions, including the quartic coupling $\lambda$, into concordance with
experiment, virtually eliminating fine-tuning and predicting the new mass scale
of the binding interaction, $M_0\sim 6$ TeV,  ref.(\cite{chris1},\cite{chris2}) summarized in Appendix B. 

We emphasize  that
the theory is  {\em manifestly} Lorentz invariant.
The challenge is, however, that the low energy internal wave-function, $\phi(r^\mu)$, introduces {\em a priori} unwanted dependence
upon  ``relative time.''  This is $r^0 =(x^0-y^0)/2 $ in the rest frame of the bound state, but boosted relative time would occur
in any frame.  Due to the single time parameter of Hamiltonian based quantum mechanics, the relative time is unphysical 
in a wave-function.  

The wave-function can be viewed as the ``end-cap'' of the path integral that begins (and terminates) on  given time-slices. The path-integral is a Green's function of the Hamiltonian--Schr\"odinger equation that propagates
the wave-function from an initial time and spatial configuration  $(t_0, \vec{X}_0,\vec{r})$ to
a future $(t_1, \vec{X}_1,\vec{r}_1)$ (this is true in field theory where the initial and
final configurations are {\em static fields}, $\phi(\vec{x}_i)$, specified on given time slices, $t$).  The relative time does not exist on these end-cap time slices and plays no role in the initial data.\footnote{
We remind the reader that, for  the 
Hydrogen atom with potential, $V(\vec{r})$ and Schr\"odinger equation 
$H\psi(\vec{r},t)=i\partial_t\psi(\vec{r},t)$, the
parameters $t$ and $\vec{r}$ {\em do not form a 4-vector} under Lorentz covariance. Rather, 
$(t,\vec{X})\rightarrow X^\mu$ becomes a relativistic 4-vector, with $\vec{X}$ describing the center-of-mass of the atom, 
and $\vec{r}\rightarrow r^\mu$ with the relative time $r^0$ constrained to zero.}
Hence, for wave-functions, relative time must be removed in a manner consistent with Lorentz invariance. 
The issue of relative time is avoided in the NJL model due to the pointlike interaction, 
 but the absence of relative time is a well known  challenge  known to arise in 
 any bound state with an extended interaction, \cite{Dirac}. 

 The problem of maintaining Lorentz invariance is that relative time implies
 an ``arrow of relative time,'' a timelike, unit, 4-vector $\omega^\mu$,
 associated
 with $\phi(r^\mu)$.  The relative time is then $\tau$, where $r^\mu =\omega^\mu \tau$.   
The absence of relative time can then be viewed as a gauge symmetry of the internal wave-function $\phi(r^\mu)$,
where a gauge transformation is  $r^\mu \rightarrow r^\mu + \omega^\mu \tau$
with $\tau$ acting as a gauge parameter.
Given an $\omega^\mu$
we can  then pass to a manifestly gauge invariant $\phi_\omega(r^\mu)$ field, the analogue of a ``Stueckelberg'' field,  where the symmetry 
is built in:
\bbo
\label{eight}
\phi(r^\mu) \rightarrow \phi_\omega(r^\mu) \equiv \phi\left( \omega^\mu \omega_\nu r^\nu-r^\mu\right) \qquad 
\omega^2 \equiv \omega_\mu\omega^\mu =1.
\ebo 

The issue then becomes ``what defines $\omega^\mu$?''
For simple two-body bound states, where  $H(X^\mu)\sim \exp(iP_\mu X^\mu)$, 
then $\omega^\mu$ can be identified with the normalized 4-momentum, $\omega^\mu =P^\mu/\sqrt{P^2}$,
where the bound state mass is  $P^2 = P_\mu P^\mu = \mu^2 >  0$.
Hence, a two-body spherically symmetrical  bound state becomes:
 \bbo
 \label{soln}
 H(X^\mu)\phi_\omega(r^\mu) = \exp(iP_\mu X^\mu)\; \phi\!\left(\sqrt{((P_\mu r^\mu)^2/P^2) -r_\mu r^\mu  }\right).
 \ebo
 This state is then explicitly Lorentz invariant. Then
 $\phi(r^\mu)$  reverts to $\phi(|\vvr|)$
in the rest frame where it can be treated as a solution to a static Schr\"odinger-Klein-Gordon (SKG) equation \cite{chris1}.

 However, there remains the  question  when $\mu^2 < 0$ and spontaneous symmetry breaking occurs:
``How can we  have a Lorentz invariant  vacuum state with $P^\mu=0$, but nonzero $\omega^\mu$?''  That is, ``what determines $\omega^\mu$ in the vacuum?''
Any constraint that locks $P^\mu$ to $\omega^\mu$
ceases to exist in the vacuum and $\omega^\mu$ would seem to become arbitrary. 
One possibility is that $\phi(r^\mu)\rightarrow $(constant) in the vacuum, but
this does not lead to a consistent solution to the SKG equation with a potential that
depends upon $r^\mu$.  

It would seem nonsensical to assert that the vacuum is defined by a 
condensate that occurred in a particular Lorentz frame with a random $\omega^\mu$. 
If, for example,  $\omega^\mu$ is somehow associated with the 
local cosmic rest frame we would obtain induced Lorentz violating effects in the electroweak physics, e.g., vacuum Cerenkov radiation for all particles that receive mass from the BEH boson, top-quarks to electrons and neutrinos \cite{ColemanGlashow}. 
We have estimated these effects in \cite{chris1} and, for a particle of mass $m_f$, they are suppressed as $\sim  m_f|\mu|^2/M_0^2$. While the limit
on vacuum Cerenkov radiation for the electron is satisfied, these effects are likely problematic,
e.g., potentially large
radiatively induced Lorentz non-invariant corrections to electrodynamics 
may arise at loop level \cite{Kostelecky}.
We therefore require a starting point 
in which the vacuum is manifestly Lorentz invariant, hence it must contain no preferred $\omega^\mu$.
Yet, we evidently require nonzero
 $\omega^\mu$ to define the solutions, $\phi_\omega(r^\mu)$, to implement the relative time invariance.

 Hence, we arrive at the following solution to the vacuum problem:
{\em the vacuum is a Lorentz invariant sum over all frames of the individual solutions $\phi_\omega(r^\mu)$ in each frame.} 
We introduce a Lorentz invariant integral over $\omega^\mu$,
leading to a novel  internal wave-function for the vacuum state, $\Phi(r^\mu)$,
with (unrenormalized) $H'(X)$.
\bbo
\label{defdef}
  H(x,y) = H'(X)\Phi(r^\mu) \qquad \text{where,} \qquad \Phi(r^\mu)=
  {\cal{N}}\int d^4\omega \;\delta(\omega^2-1) \;\phi_\omega(r^\mu).
\ebo
With this definition of the vacuum,  $\Phi(r^\mu)$ becomes  a collective state,  similar
to a condensed matter system such as the BCS superconductor.  
Note the return of relative time, $r^0$, but we will see that $r^\mu$ can now be integrated
out of the final action, yielding Lorentz invariant higher dimension operators
suppressed as $\sim M_0^{-2p}$ for $p\geq 1$.

Our present calculation is formally similar to the construction of the coherent condensate of Cooper pairs
in a BCS superconductor  \cite{Cooper}\cite{BCS}, differing essentially by
our required Lorentz invariance.  Cooper pairs are
two-body bound states, (approximate) bilocal plane waves  
in momentum space: $\phi_{i}(\vec{r})\approx \phi_i\exp(\vec{k}_i\cdot(\vec{x}-\vec{y}))  $,
where the $\vec{k}_i$ lie in a small range immediately above the Fermi surface of the material,
hence the ``pairing'' is an electron of momentum $\vec{k}_i$ with the antipodal electron $-\vec{k}_i$.
The Hamiltonian kinetic term for the pairs is then $\small \sim (1/m)\sum_{i} |\phi_{k_i}|^2 k^2_i \propto N $,
while the  two-body attractive scattering potential (weak phonon exchange interaction) is 
$\small \sim - \epsilon m \sum_{i} \sum_{j} \phi^\dagger_{k_i}\phi_{k_j}\propto N^2  $. Hence
the weak phonon  interaction is significantly enhanced by large $N\epsilon \sim 1$ and can compete
with the kinetic terms to form a stable collective state
which forms the condensate. The semiclassical wave-function of the ground state is then 
$\Phi \sim \sum_i^N \phi_i(\vec{r})$ where the sum spans the Fermi surface.  The BCS theory is 
essentially a bilocal field theory.\footnote{
The BCS ground state is a coherent state of Cooper pairs that is a product of Dirac kets
that are mixtures of the Fermi-degenerate groundstate and pairs, 
$|BCS\rangle \sim\prod^{N_F}_i (u_{i}|0\rangle + v_{i}|\!\uparrow\!\vec{k}_i, \!\downarrow\! -\vec{k}_i\rangle)$,
where the product extends over the $N_F$ momenta, $k_i$, slightly
above the Fermi surface (bounded above by the Debye frequency $\omega_D$).
The  superconductor's
properties are determined by  the resulting ratio  $v_{i}/u_{i}$, obtained by
minimization of the Hamiltonian with, e.g.,  temperature effects
or current dependence,  etc. (See e.g.,
 \cite{Feynstat}\cite{Urbana}, and related discussion of mean-field approximations  \cite{GL} \cite{BEC}). 
 The  condensate quantum field is bilocal, with creation operators $a^\dagger_{\vec{k}_i}$ for electrons of momenta
 $\vec{k}_i$, 
 $\widehat\Phi(\vec{r}) \sim  \sum^N_i a^\dagger_{\vec{k}_i} a^\dagger_{-{\vec{k}_i}}\phi_{i} (\vec{r}) $. Note
 this creates, and sums over, Cooper pairs of antipodal momenta $(\vec{k}_i, -\vec{k}_i)$ that span the Fermi surface. The analogue of our semiclassical bilocal
 field is then $\Phi(\vec{r}) = \langle BCS|\widehat\Phi(\vec{r})|BCS\rangle \sim
 \sum^N_i u_iv_i^*\phi_{i}(\vec{r}) $. The ground state (vacuum)  expectation value of $\langle \Phi \rangle$ is determined by
 a gap equation (similar of the original formulation of the NJL model \cite{NJL}). A gap
 equation is equivalent to the minimization of the BEH (Higgs) potential (the ``quartic interaction'' of the potential
 is implicit in the gap equation loop, and for us the quartic coupling also arises at loop level).
} 

The Brout-Englert-Higgs (BEH) vacuum wave function which we presently propose,  $\Phi(r^\mu)$,
is a sum over all {\em constituent wave-function solutions of the SKG equation}, each with $P^\mu =0$ 
but with nontrivial  $\phi_\omega(r^\mu)$,  each
corresponding to a different Lorentz frame, $\omega^\mu$, the analogue of the Fermi momenta in the superconductor.
The  $\phi_\omega(r^\mu)$ are then a 
relativistic generalization of Cooper pairs  \cite{BCS}, but while Cooper pairs  span
the Fermi surface in a superconducting condensate, the $\Phi \sim \int_\omega \phi_\omega$
span the future timelike hyperboloid in the timelike unit 4-vectors, $\omega^\mu$.  

The component fields, $\phi_\omega(r)$, each satisfy a nontrivial {\em integro-differential equation }
with  dependence upon the frame $\omega^\mu$.
It follows that the invariant integral over 
$\omega^\mu$, normalized by   ${\cal{N}}$,
yields a Lorentz invariant  $\Phi(r^\mu)$.
The Hamiltonian is then  diagonalized by the coherent state $H(X)\Phi(r^\mu)$.  
The normalized effective action for $H(X^\mu)$ 
is obtained and, upon integrating out $r^\mu$, yielding the ``Higgs'' potential $\mu^2|H|^2 + (\lambda/2)|H|^4$.
$H$ then acquires  a vacuum expectation value (VEV),  determined by a negative SKG eigenvalue, $\mu^2$, 
and stabilized by the loop induced quartic interaction $\lambda$. 

\begin{figure}
	\centering
	\hspace{-0.7in}
	\includegraphics[width=0.6\textwidth]{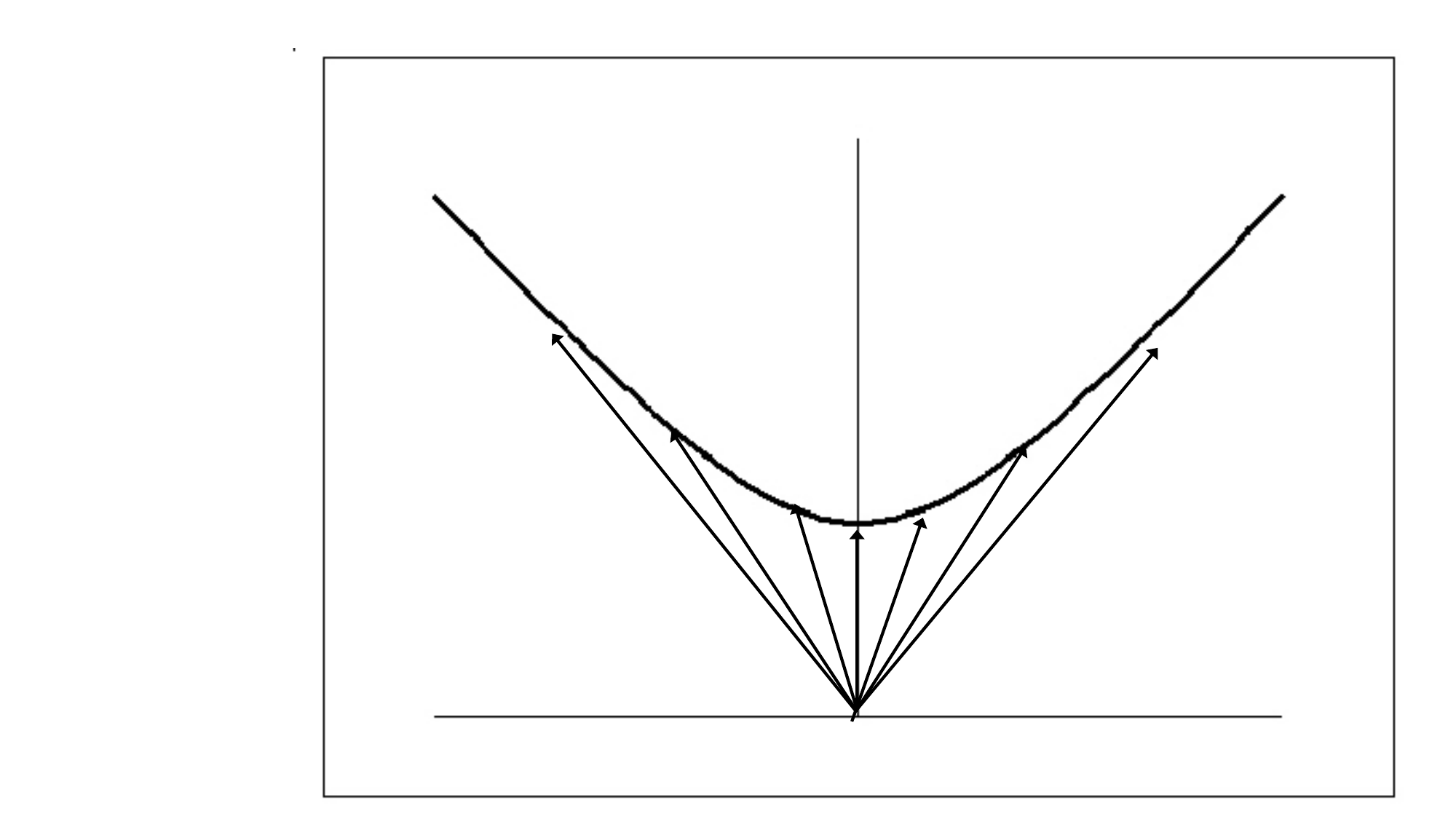}
	\caption{ \textmd{\small Vacuum wave-function, $\Phi(r^\mu)$, spanning the timelike hyperboloid in 4-vectors $\omega^\mu$ by integrating over ``internal wave-functions,'' $\phi_\omega(r)$, to form the Lorentz invariant $ \Phi(r^\mu)={\cal{N}}\int d^4\omega \;\delta(\omega^2-1) \;\phi_\omega(r^\mu)$. This is a relativistic analogue of a BCS state that integrates over the Cooper pairs on the Fermi surface.}} 
	\label{fig:loop}
\end{figure}

 The vacuum emerges from the underlying theory  
as $\langle H\rangle {\Phi}(r)$ where $ \Phi(r^\mu)$
is the coherent sum over all $\phi_\omega(r^\mu)$ and $\langle H\rangle =v_{weak}$.
The BEH boson observed  at the LHC, $h(X)$ and 
the Nambu--Goldstone phases that become longitudinal $W^\pm$ and $Z^0$, then
emerge
as  ``excitons'' of the collective state.
In the broken phase we then have:
  \bbo
 H(X^\mu){\Phi}(r^\mu)  \rightarrow \exp(i \pi^a(X)\tau^a/2v_{weak}) \bl\begin{array}{c} v_{weak} + \frac{1}{\sqrt{2}}h(X) \\ 0 \end{array}\br{\Phi}(r^\mu),
\ebo
where $h(X)$ is the ``Higgs field'' of the standard model.

Here the constant zero 4-momentum VEV, $v_{weak}$, is carried by $H(X^\mu)$ and determined in the usual 
way by the minimum of the sombrero potential. 
The main prediction of the theory remains as the existence of 
the new binding interaction at the scale $M_0\approx 6$ TeV and the emergence of
an octet of colorons coupled most strongly to third generation quarks.  This interaction is only partially strong (approximately half critical) since the critical behavior
occurs only in the binding channel where loop effects reinforce the binding interaction \cite{chris2}\cite{Jackiw}.

 In the  present
paper we will mainly focus on
the nontrivial vacuum. We will also briefly sketch how higher dimension operator terms ${\cal{O}}(1/M_0^2)$
may be extracted from the Yukawa coupling
in Section IV (see also \cite{chris1}), which may ultimately present observables in
high sensitivity flavor physics experiments.
We mainly ignore the complications of the introduction of all flavors of quarks and leptons,
in particular the $b_R$ quark. We expect these to be  perturbative and
follow the earlier papers on extended technicolor \cite{Technicolor}
and ``topcolor'' from the 1990's \cite{Topcolor}, and will be the subject of
future work \cite{christodo}.

Our main result is a successful,  natural,   minimally fine-tuned (few $\%$),
composite theory of the BEH boson,
with a novel physical scale $\sim 6$ TeV corresponding to a new semi-strong interaction
within the third generation quarks, and associated gauge fields that are possibly accessible to the LHC
(\cite{chris1},\cite{chris2}). We show that vacuum of this theory maintains the Lorentz invariance,
as a dynamical analogue of a BCS superconductor, yielding the spontaneous symmetry breaking
seen in the SM and a composite BEH boson. 

\newpage

\section{Implementing the  Relative Time Symmetry }

We begin with the notion that
the internal wave-function of two-body bound state, $\phi(r^\mu)$, must not
 depend upon ``relative time,'' i.e., is independent of $r^0$ in the barycentric frame, or
 correspondingly $r'^0$ in any frame. This can be seen in the free field limit by kinematics if we
consider a pair of equal-mass particles of 4-momenta $p_1$ and $p_2$, $p_1^2=p_2^2=m^2$,
and a bilocal wave-function,
$H(x,y) \sim \exp(ip_1x+ip_2y)$.  We  pass to the total
momentum $P=(p_1+p_2)$ and relative momentum  $Q=(p_1-p_2)$,
and the plane waves become $\exp(iPX+iQr)$, where we define ``barycentric coordinates,''
$x^\mu=X^\mu+r^\mu$,  $y^\mu=X^\mu-r^\mu$.
Note that $P_\mu Q^\mu=p_1^2-p_2^2=0$. Therefore,
in the center-of-mass frame, in which
$P=(P^0,\vec{0})$ and $Q=(0,\vvq)$, we see that  $Q^0=0$. This implies there is  no dependence
in the bilocal state on $r^0$ through $\exp(iQ_0r^0)=1$, and likewise no dependence upon
a boosted relative time $r'^0$ in any other frame.  
If the particles are constituents of a bound state, then  the ``relative time'' 
must decouple from the dynamics. Indeed, it is useful to think of the bound state as free field
pair of particles for which the bilocal field is properly normalized, and the interaction is subsequently
adiabatically switched on, maintaining this kinematic constraint.  The free particle states then ``flow'' to become the bound states.

Given
an arbitrary Lorentz invariant function, $\phi(r^\mu)$,  in any frame there will generally be
dependence upon  a relative time, $\sim \tau$. This is
analogous to the gauge dependent components of a vector potential, and it  is an artifact of using the bilocal field description.\footnote{I am grateful to Bill Bardeen for 
some discussions that inspired this perspective.}    As described
in the Introduction, the relative time, $\tau$, can be written in any given frame as $r^\mu = \omega^\mu \tau$, and
implicitly requires the timelike, 4-vector, $\omega^\mu$, the ``arrow of relative time.'' 
Hence,  eliminating dependence upon $\tau$, requires a Lorentz invariant
constraint, such as  $\omega_\mu\partial^\mu \phi(r)=0 $.

In the symmetric phase of the standard model (SM), (or for any typical  two-body
bound state) the BEH boson contains such a vector, i.e., the 4-momentum $P^\mu$ carried by 
$H(X^\mu)\sim \exp(iP_\mu X^\mu)$.  We can therefore
bootstrap $\omega^\mu$ to $P^\mu$ through a constraint relation
$\omega^\mu\propto P^\mu$. For example, we can do this semiclassically by introducing Lagrange multipliers  
into the action, such as:
\bbo
W= M_0^4 \int d^4X d^4r \;\lambda'(H^\dagger D_\mu H \phi^\dagger \partial_\mu \phi),
\ebo
where we demand $\delta W/\delta\lambda' = 0$,
which imposes the kinematic constraint  $P_\mu Q^\mu=0$.
However, the question then remains: ``what happens in the vacuum  where  $P_\mu =0$
and $\omega^\mu$ becomes unconstrained?''

Consider the bilocal field in barycentric coordinates:
\bbo
H(x,y) \rightarrow H(X^\mu)\phi_\omega(r^\mu).
\ebo
We will presently focus upon the fields, $\phi_\omega(r^\mu) $, as defined in eq.(\ref{eight})  which 
are invariant under the gauge transformation $r^\mu\rightarrow \omega^\mu \tau$  and thus have
no dependence upon  $ \tau$, though an implicit dependence upon $\omega^\mu$ remains.\footnote{Indeed, $\phi_\omega(r^\mu)  $ is
then  the analogue of
a ``Stueckelberg'' field, e.g., a gauge field such as $B_\mu =A_\mu - \partial_\mu \chi$
which is invariant under $ A_\mu \rightarrow A_\mu+ \partial_\mu \tau$ and $\chi\rightarrow \chi+\tau$. } 
Note that eq.(\ref{eight}) satisfies
the  constraint equation  $0=\omega^\mu \partial_\mu \phi(r^\mu)$.  
In the following, $\phi(r^\mu)$  will refer to an arbitrary Lorentz invariant scalar, while $\phi_\omega(r^\mu)$
is of the (Stueckelberg) form of eq.(\ref{eight}) with the relative time projected out.

Consider the action, $S_\phi$ of eq.(\ref{bcs}), for the internal field $\phi(r^\mu)$ that arises in natural top condensation \cite{chris1}. 
We  replace $\phi(r^\mu)\rightarrow \phi_\omega(r^\mu)$ in  $S_\phi$,
 \bea 
 \label{bcs2}
\backo 
S_\phi & \rightarrow  & M_0^4\int \!\! d^4r \bl Z \partial_\mu\phi_\omega^\dagger(r) \partial^\mu \phi_\omega(r) 
+ 2g_0^2N_c D_F(2r^\mu) \phi_\omega^\dagger(r)\phi_\omega(r)\br,
\eea
where $\partial_\mu = \partial/\partial r^\mu$  and $D_F(2r) $ is the potential as defined in eq.(\ref{DF}).

If we then vary  $S_\phi$ with respect to $\phi_\omega+\delta\phi_\omega$ we obtain a formal, manifestly Lorentz invariant 
{\em integro-differential equation}:
\bbo
\label{nine}
M_0\int\!\! \bl - Z  \frac{\partial^2 \phi_\omega(r^\mu)}{\partial r^\mu \partial r_\mu}
+ 2g_0^2 N_c D_F(2r^\mu) \phi_\omega(r^\mu)\br\omega_\nu dr^\nu = ZM_0\int \!\! \mu^2\phi_\omega(r^\mu)\omega_\nu dr^\nu = \mu^2\phi_\omega(r^\mu).
\ebo
Note  the presence of the overall line integral, $\int \omega_\mu dr^\mu$. 
This line integral remains in the equation of motion since $\phi_\omega(r^\mu)$ has no dependence
upon $r_\mu \propto \omega_\mu $, hence the variation is constrained, $\delta\phi_\omega  \sim \delta^3(r^\mu_\perp)$ 
where
$\omega_\mu r^\mu_\perp = 0$ and does not produce a longitudinal variation, 
$\delta(\omega_\mu r^\mu$). It is important to realize that eq.(\ref{nine}) is not a conventional
Klein-Gordon equation due to the line integral constraint.
Here we define $Z$ by the line integral normalization \cite{chris1}, 
\bbo
\label{line}
Z M_0 \int dr^\mu \omega_\mu  =1.
\ebo
$Z$ is thus defined to canonically normalize the free field pair of particles before the interaction
is turned on, where $\phi_\omega$ is a dimensionless field and normalized as:
\bbo
\label{normn}
M_0^4 Z\int \!\! d^4r |\phi(r^\mu)|^2 =1.
\ebo

We can find solutions to eq.(\ref{nine}) as follows:
Since $S_\phi$ is Lorentz invariant
we can evaluate the action in the particular frame, $\omega^\mu =(1,0,0,0)$,
where $\phi_\omega(r^\mu) \rightarrow \phi(0, \vec{r})\equiv \phi(\vec{r})$
and hence  $Z M_0\int dr^\mu \omega_\mu \rightarrow Z M_0\int dr^0 =1$. 
In this  frame the action becomes:
 \bbo
  \label{intD}
\backo S_\phi = M_0^3\!\!\int \!\! d^3r\; \bl
-|\nabla_{\vvr}\phi(\vvr)|^2 
+\int dr^0  \;{2g_0^2 N_c M_0 }D_F(2r^\mu)|\phi(\vvr)\; |^2
\br.
 \ebo 
 Using eq.(\ref{line})
converts the normalization of eq.(\ref{normn}) to:
\bbo
\label{newnorm}
ZM_0^4\int \!\! d^4r |\phi(r^\mu)|^2 \rightarrow 
\int \!\! d^3r\; M_0^3\;|\phi(\vvr)|^2 =1,
\ebo
and yields, in this frame, the Yukawa potential \cite{chris1}:
\bbo
V(2|\vvr|) = \int dr^0 2g_0^2 N_c D_F(2r^\mu)= -\frac{g_0^2 N_c e^{-2M_0|\vvr|}}{8\pi |\vvr|}.
\ebo
The SKG equation in the spherical ground state thus becomes:
\bbo
\label{SKG}
\backo\!\!\!\!
-
\nabla^2\phi(r)-g_0^2N_c M\frac{ e^{-2M_0 r}}{8\pi r}\phi(r) =\mu^2\phi(r)
\qquad  \nabla^2=\bl\frac{\partial^2 }{\partial r^2}+\frac{2}{r}\frac{\partial }{\partial r}\br
\qquad r\equiv \sqrt{\vvr^2}.
   \ebo    
 We have discussed its properties and solutions in ref.\cite{chris1}.

 In the solution to the SKG equation with the eigenvalue, $\mu^2$, we see upon
 integrating by parts that $S_\phi\rightarrow \mu^2$, 
 and the full bound state action, given in eq.\ref{Haction2} of Appendix B, then becomes:
  \bbo
  \label{action20}
\backo\backo S=
\!\!\int\!\! d^4X\;\bl
 |D_H H(X^\mu)|^2-\mu^2 
|H(X^\mu)|^2 -\frac{\lambda}{2}(H^\dagger H)^2 - g_Y\left([\bar{\psi}_{iL}(X)t_{R}(X)]_{f}H^i(X)  +h.c.\right)\br + S'.
   \ebo   
$\mu^2$  is the physical mass  of the bound state.
Note that in the {\em pointlike limit of the interaction}, the internal fields $\phi_\omega(\vec{r})$ remain extended.
The Yukawa coupling, $g_Y$ and quartic coupling $\lambda$ are
defined in eqs.(\ref{A6}-\ref{A11})  in Appendix B.
We emphasize that, while we have evaluated the action in the particular $\omega^\mu$ 
frame in eq.(\ref{intD}), this is just a
calculation in a simplifying frame  
in the overall Lorentz invariant action eq.(\ref{nine}). The result eq.(\ref{action20}) holds in any frame.

As shown in ref.\cite{chris1}, the SKG equation has
a critical coupling, $g_0=g_c$, for which $\mu^2=0$, very close to the quantum
NJL critical coupling:
\bbo
\label{exact}
\frac{g_c^2N_c}{8\pi^2} =1.06940,\qquad \makebox{c.f, the NJL critical value, } 
\qquad \left.\frac{g_0^2N_c}{8\pi^2}\right|_{NJL}=1.00.
 \ebo
 The loop level (NJL-like) effects generate $\lambda$ and also
 add to the formation of the bound state as discussed in \cite{chris2}.
 This amplifies the coupling strength in the bound state channel and generates a  renormalized coupling
for the  4-fermion interaction,
$\bar{g}^2_0$, where:
\bbo
\bar{g}_0^2 = g_0^2\left(1-\frac{g_0^2N_c}{8\pi^2} \right)^{-1}.
\ebo
When $\bar{g}_0>g_c$ the eigenvalue is $\mu^2<0$.
In such a solution the action eq.(\ref{action20})
for $H(X^\mu)$ then yields the ``sombrero potential'':
 \bbo
 \label{sombrero}
\mu^2 |H|^2 +\frac{\lambda}{2}(H^\dagger H)^2 \qquad \text{where,} \;\;\; \mu^2 <0.
   \ebo   
  The solution of the SKG equation for $\phi(r)$ can be obtained approximately analytically, or
  by numerical integration \cite{chris1, chris2}.
  At short distances $\phi(r)\sim \phi(0)$, and extends at large distances in the rest frame
  to, $\phi(r) \sim ce^{-|\mu|r}/r$, where $|\mu|<\!\!M_0$ and we are near critical coupling.
  The solution is normalized as in  eq.(\ref{newnorm}), which 
  dilutes the value of $\phi(0) \sim \sqrt{|\mu|/M_0}$ and suppresses 
  the Yukawa coupling $g_Y \propto \phi(0)$. Inputting $g_Y\approx 1$ yields $M_0\approx 6$ TeV, and  we find
  $\lambda\propto g_Y^4 $, as given by loops  in eq.(\ref{A11}, remarkably close to the expermental value.

  \section{Broken Phase, Manifestly Lorentz Invariant Vacuum, and BEH Excitations}

  Presently we show how the action can be written in terms of the collective field $\Phi(r^\mu)$.
  We find it conceptually useful to begin by  approximating an integral representation
of the collective state by
 a discrete sum over $N$ of the solutions, $\omega_{i\mu}$. This approximates the  continuous integrals which are
 defined subsequently corresponding to the  large-$N$ limit.   We begin with a brief discussion of the orthogonality
 of the $\phi_{\omega_i}$ which is further treated in Appendix A.  In Appendix B for reference, we give
 a formal summary of the natural top condensation scheme.

  \subsection{Formal Derivation}
  
 The
 bound state action of natural top condensation in eq.(40) of ref.\cite{chris1} and takes the form: 
  \bbo
\label{Haction0}
\backo\backo
S=
 M_0^4\!\!\int\!\!d^4X\;d^4r \bl Z|D_{H}H(X^\mu) |^2|\phi(r) |^2
+Z|H(X^\mu)|^2|\partial_{r}\phi(r) |^2 +2{g_0^2 N_c}D_F(2r)|H^\dagger H||\phi(r)) |^2\br \!\! +S_Y+S_\lambda + ...
 \ebo   
where the composite field is $H(X^\mu)\phi(r)$. The $H$ kinetic term is canonical with 
the Lorentz invariant normalization of eq.(\ref{normn}) and
$Z$ is as defined upon removal of relative time as in eq.(\ref{line}). The Yukawa interaction is generated at tree level,
\bbo 
\label{SY}
\backo
S_Y 
= \widehat{g}_Y M_0^2
\int\!\! d^4X d^4r \;[\bar{\psi}_{iL}(X\!+\!r)\psi_{R}(X\!-\!r)]_{f}D_F(2r)\;H^i(X)\phi(\vvr) {+h.c.},
\ebo
and a quartic interaction is generated at loop level, given in the point-like $\phi(r)$ approximation by:
\bbo 
\label{Slam2}
\backo
S_\lambda 
=  -\frac{\widehat{\lambda}}{2}
\int\!\! d^4X  (H^\dagger H)^2 |\phi(0)|^4 {+h.c.}. = -\frac{\lambda}{2}\int\!\! d^4X  (H^\dagger H)^2  {+h.c.}.
\ebo
 In the above, the Yukawa and quartic couplings  $g_Y$ and $\lambda$ are derived quantities 
from the underlying theory. In a pointlike approximation for the interaction, $D(2r)\sim \delta^4(r)/JM_0^2$, 
(where $J=16$ is the Jacobian for $(x,y)\rightarrow (X,r)$) and  we have (from eq.(77) of ref.\cite{chris1} ):
\bbo
\label{A6}
\widehat{g}_Y\approx  g_0^2 \sqrt{2N_c/J}\;\phi(0),
\qquad\qquad
\lambda \approx (g_Y^{4}-g_Y^2\lambda) \frac{N_{c}}{4\pi ^{2}}%
 \ln \left( \frac{M_0}{\mu }\right) .
\ebo
Note that $\widehat{g}_Y$ is classical and  $\lambda$ arises at loop level (${\cal{O}}(\hbar)$). 
We thus see that these are subject to dilution in an extended solution with the internal wave-function  $\phi(r)$
and $g_Y= \widehat{g}_Y \phi(0)$ and $\phi(0)\propto \sqrt{|\mu|/M_0}$.
Experimentally, with $|\mu|=88$ GeV, and inputting $g_Y \approx 1$ we determine $M_0\approx 6$ TeV, 
and  $\lambda \approx 0.23$.

 We now consider $\phi(r) \rightarrow \phi_\omega(r)$, the ``Stueckelberg'' fields of eq.(\ref{eight}) 
  with relative time projected out.  Our strategy is to replace, in the action, the composite field $H(X)\phi(r)$
  considered in ref.\cite{chris1}, by the new field $H'(X) \Phi(r)$, where $\Phi(r)$ is the coherent sum over $\phi_\omega(r)$.
  With suitable normaliziation conventions
  we will show that this reduces back to the standard model BEH action with the Lorentz invariant corrections
  of order $1/M_0^2$.

We can see that there is orthogonality between the $\phi_\omega(r^\mu)$ solutions with different $\omega$'s.
Formally, for a smooth Lorentz invariant function,  $F(r^\mu)$ (or differential operator, e.g, $F\sim \partial^2$):
\bbo
\label{240}
Z M_0^4\int \!\! d^4r \;   \phi_{\omega}^\dagger(r^\mu) F(r^\mu)   \phi_{\omega`}(r^\mu)
 =  ZM_0^4\int \!\! d^4r \;
 \delta^4(\omega-\omega') \phi_{\omega}^\dagger(r^\mu) F(r^\mu)  
 \phi_{\omega`}(r^\mu).
 \ebo
If however, $F\sim\delta^4(r^\mu)$ we have mixing of $\omega$ and $\omega'$:
\bbo
\label{ifdelta}
Z M_0^4\int \!\! d^4r \;   \phi_{\omega}^\dagger(r^\mu) \delta^4(r^\mu)  \phi_{\omega`}(r^\mu)
=  ZM_0^4\phi_{\omega}^\dagger(0) \phi_{\omega`}(0)
 =  ZM_0^4\int \!\! d^4r \;\delta^4(r^\mu)\phi_{\omega}^\dagger(r) \phi_{\omega`}(r).
 \ebo
We derive the orthogonality in Appendix A.

For transparency, we will presently form a coherent state by a discrete sum over fields $\phi_{\omega_i}$.
\bbo
\backo
{\cal{N}}\int d^4\omega \delta(\omega^2-1) F(\phi_\omega) \;\;  \longrightarrow \;\; \sum^N_{i=1} F(\Phi_{\omega_i}).
\ebo
The $\phi_{\omega_i}(r)$ each have an internal $\omega_{i\mu}$ and each is a solution
of the equation of motion, eq.(\ref{nine}). 
The orthogonality of the $\phi_{\omega_i} $ fields implies in a discrete sum:
\bbo
\label{31}
\int \!\! d^4r \;\phi_{\omega_i}^\dagger(r)\phi_{\omega_j}(r)
=\delta_{ij} \int \!\! d^4r \;\phi_{\omega}^\dagger(r)\phi_{\omega}(r),
\nonumbo
\int \!\! d^4r \; \partial_\mu\phi_{\omega_i}^\dagger(r)\partial^\mu \phi_{\omega_j}(r)
=\delta_{ij} \int \!\! d^4r 
\;\partial_\mu\phi_{\omega}^\dagger(r) \partial^\mu \phi_{\omega}(r) .
\ebo

Therefore, we define the collective field by summing over a large set of $N$  arbitrary $\omega_{i\mu}$ unit 4-vectors that 
span the future timelike hyperboloid:
  \bbo
  \label{newfield}
\Phi(r^\mu) = \frac{1}{\sqrt{N}} \sum^N_{i=1} \phi_{\omega_i}(r^\mu), \qquad \text{where,}\qquad  \phi_{\omega_i}(r^\mu) = \phi(\omega_i^\mu \omega_i^{\nu} r_\nu -r^\mu), \qquad \omega_i^2=1.
\ebo
In this framework we will also have the classical average of the $\phi_\omega$:
\bbo
\widetilde{\Phi}(r^\mu) = \frac{1}{\sqrt{N}} \Phi(r^\mu)= \frac{1}{{N}} \sum^N_{i=1} \phi_{\omega_i}(r^\mu);
\qquad \text{Note that for any $\omega_i$:}\qquad  \widetilde{\Phi}(0)=\phi_{\omega_i}(0).
\ebo

Consider
the action for the single two-body bound state $H(X^\mu)\phi(r)$,  
replacing $\phi(r)\rightarrow \Phi(r^\mu)$ and $H(x)\rightarrow H'(X)$.
and denote renormalized  parameters by primes $'$.  For the sake of discussion we break 
the action into separate components:
\bbo
\label{Haction}
S= S_1 + S_2 + S_3 +S_Y+S_\lambda, \qquad \text{where we define,}\qquad H(X^\mu)\phi(r)\rightarrow H'(X)\Phi(r^\mu),
 \ebo   
and:
 \bbo\label{S1}
 S_1=  M_0^4\!\!\int\!\!d^4X\;d^4r \bl Z'|D H'(X) |^2|\Phi(r^\mu) |^2\br,
 \ebo
 \bbo\label{S2}
 S_2=M_0^4\!\!\int\!\!d^4X\;d^4r \bl Z'|H'(X)|^2|\partial_{r}\Phi(r^\mu) |^2\br ,
 \ebo
  \bbo\label{S3}
 S_3=M_0^4\!\!\int\!\!d^4X\;d^4r \bl 2{g'{}_0^2 N_c}\frac{1}{16M_0^2}\delta^4(r^\mu)|H'^\dagger H'||\Phi(r^\mu)) |^2\br ,
 \ebo
where we have here taken the pointlike limit of $D_F(2r^\mu)$ of eq.(\ref{DF}),
\bbo
D_F(2r^\mu)\rightarrow  \frac{1}{M_0^2}\delta^4(2r^\mu) = \frac{1}{16M_0^2}\delta^4(r^\mu). 
 \ebo
 The Yukawa interaction is,
\bbo 
\label{SY2}
\backo
S_Y 
= \widehat{g}_Y M_0^2
\int\!\! d^4X d^4r \;[\bar{\psi}_{iL}(X\!+\!r)\psi_{R}(X\!-\!r)]_{f}D_F(2r)\;H'^i(X)\Phi(\vvr) {+h.c.},
\ebo
 and the quartic interaction,:
\bbo 
\label{Slam}
\backo
S_\lambda 
=  -\frac{\widehat{\lambda}}{2}
\int\!\! d^4X  (H'^\dagger H')^2 |\Phi(0)|^4 {+h.c.}. 
\ebo
Our problem is to verify that $S(H'\Phi)$, with renormalized parameters, is  consistent with the underlying
theory of the $\phi_\omega=\phi(r)$ as defined in eqs.(\ref{Haction},\ref{bcs}). We therefore
substitute the
collective field definition of eq.\ref{newfield} and define the renormalized parameters.

We define the following normalizations:

\begin{centering}
  \bbo 
  \begin{array}{|cccccc|}
  \hline 
   & & & & & \\ 
  & \qquad  Z' = NZ ,\qquad 
  & \qquad H'= \frac{1}{\sqrt{N}} H, \qquad
  &  \qquad \Phi(r^\mu) = \frac{1}{\sqrt{N}} \sum^N_{i=1} \phi_{\omega_i}(r^\mu)  ,
  &  \qquad \widetilde{\Phi}(r)   =\frac{1}{\sqrt{N}}\Phi(r^\mu).\qquad & \\
  & & & & & \\ \hline
  \end{array}  \nonumber
  \ebo
\end{centering}

\noindent
Note that these normalizations differ from the BCS superconductor in two major regards. 
First, $Z$ is treated as an extensive parameter, i.e., when 
we renormalize each $\phi_\omega\rightarrow (1/\sqrt{N})\phi_\omega$ to comprise the sum over $\omega^\mu$
 we then redefine $Z\rightarrow Z'= ZN$.  
The new composite bilocal field then takes the form $H'(X)\Phi(r^\mu)$. 

Secondly, the $H'(X)$ kinetic term should be canonical with 
the Lorentz invariant  $\Phi(r^\mu)$, hence
we renormalize the  $H'(X)$
field as:
\bbo \label{32}
H'=H/\sqrt{N}, \qquad \text{and then,}\qquad Z'= {N} Z.
\ebo
This choice preserves the product $Z' H'^\dagger H' = Z H^\dagger H$,
and $Z$ is still defined  by the line integral relation of eq.(\ref{line})
for the underlying $\phi_\omega$ field.
These normalizations will yield the same predictions for $g^2_{critical}$ and $M_0\sim 6$ TeV,
as one would obtain for a single $\phi_\omega$ condensate.

We now discuss the calculation of the full action term by term:

\vspace{0.1in}
\noindent  {\bf $S_1$: }  
We have from  eqs.(\ref{newfield},\ref{S1}):
\bbo
\label{Zr}
\backo\backo S_1=M_0^4\!\!\int\!\!d^4X\;d^4r \bl Z'|D H'(X) |^2|\Phi(r^\mu) |^2\br =
M_0^4\!\!\int\!\!d^4X\;d^4r
|D H'|^2  \bl \frac{Z'}{N} \sum^N_{i=1} \sum^N_{j=1}\phi_{\omega_i}^\dagger(r)  \phi_{\omega_j}(r)\br
\nonumbo
=M_0^4\!\!\int\!\!d^4X\;d^4r|D H'|^2  \bl\frac{Z'}{N} \sum^N_{i=1} \phi_{\omega_i}^\dagger(r)\phi_{\omega_i}(r) \br
=ZM_0^4\!\!\int\!\!d^4X\;d^4r |D H|^2   \bl  \phi_{\omega}^\dagger(r)\phi_{\omega}(r) \br\; \text{(for any }\omega) ,
\nonumbo
  = \!\!\int\!\!d^4X|D H|^2,
\ebo 
where  we use the normalization of the $\phi_{\omega_i}$ as in eq.(\ref{newnorm}).

\vspace{0.1in}
\noindent  {\bf $S_2$: }  
Likewise the $\Phi$ kinetic term becomes:
\bbo
\backo 
S_2=M_0^4\!\!\int\!\!d^4X\;d^4r \bl Z'|H'(X)|^2|\partial_{r}\Phi(r^\mu) |^2\br  
=Z M_0^4\!\!\int\!\!d^4X\;d^4r  \bl |H|^2  \partial_\mu\phi_{\omega}^\dagger(r) \partial^\mu \phi_{\omega}(r)  \br
\;\;\;(\text{for any }\omega), 
\ebo
yielding consistency with the underlying kinetic term  of the action eq.(\ref{Haction0}).

\vspace{0.1in}
\noindent  {\bf $S_3$: }  
The interaction becomes:
\bbo
 S_3=M_0^4\!\!\int\!\!d^4X\;d^4r \bl 2{g{}_0^2 N_c}D_F(2r^\mu) |H'|^2|\Phi(r^\mu)) |^2\br 
\approx M_0^4\!\!\int\!\!d^4X\;d^4r\;{|H'|^2} \bl 2g{}_0^2N_c \frac{\delta^4(r^\mu)}{16M_0^2}\frac{1}{N} \sum^N_{i=1}\phi_{\omega_i}^\dagger(r) 
\sum^N_{j=1}\phi_{\omega_j}(r) \br
\nonumbo
=M_0^4\!\!\int\!\!d^4X\;d^4r\;N|H'|^2\bl\frac{g{}_0^2N_c}{8M_0^2}\delta^4(r) \phi^\dagger_{\omega}(0) \phi_{\omega}(0)\br
=M_0^4\!\!\int\!\!d^4X\;d^4r\;|H|^2\bl\frac{g{}_0^2N_c}{8M_0^2}\delta^4(r)  \phi_{\omega}^\dagger(r) \phi_{\omega}(r) \br
\;(\text{for any }\omega).
\ebo
The essential result is that the 
potential is approximately  $\sim \delta^4(r^\mu)$, hence the line integral
orientation $\omega^\mu$ becomes irrelevant in the potential.
In the second-to-last term we see the usual enhancement factor, $N$, 
that would normally lead to the BCS enhancement phenomena.
However, in the last term the normalization of $H'= H/\sqrt{N}$ undoes the $N$-fold enhancement and
 the coupling $g_0^2$ is not renormalized.
The nonrenormalization of $g_0^2$ owes to the 
bilocal field theory with
the renormalization of $H'=H/\sqrt{N}$.

We can rewrite the interaction in the unprimed parameters as:
\bbo
S_{3}=M_0^4\!\!\int\!\!d^4X\;d^4r |H(X^\mu)|^2 \bl{2g_0^2N_c}D_F(2r) 
|\widetilde{\Phi}(r)|^2  \br.
\ebo
In the approximation $\sim {\cal{O}}(1/M_0^2)$ we can freely swap between $D_F(2r)\leftrightarrow \delta^4(r)/16M_0^2$,
and here we  use eq.(\ref{newfield}) where $\widetilde{\Phi}(0) =\phi_\omega(0)$ for any $\omega$.

\vspace{0.1in}
\noindent  {\bf $S_{Y}$: }  
The Yukawa interaction is:
\bbo 
\label{yukyuk}
\backo
S_Y 
= g^2_0\sqrt{2JN_c} M_0^2
\int\!\! d^4X d^4r \;[\bar{\psi}_{iL}(X\!+\!r)\psi_{R}(X\!-\!r)]_{f}D_F(2r)\;H'^i(X)\Phi(\vvr) {+h.c.}.
\nonumbo
\approx g^2_0\sqrt{2N_c/J}  M_0^2 
\int\!\! d^4X d^4r \;[\bar{\psi}_{iL}(X\!+\!r)\psi_{R}(X\!-\!r)]_{f}\;\delta^4(r)\;H^i(X)\widetilde{\Phi}(r^\mu) {+h.c.}.
\nonumbo
= \widehat{g}_Y M_0^2 
\int\!\! d^4X \;[\bar{\psi}_{iL}(X)\psi_{R}(X)]_{f}\;\;H^i(X)\widetilde{\Phi}(0) {+h.c.};
\qquad \qquad (\widehat{g}_Y = g_0^2 \sqrt{2N_c/J}),
\ebo
where we have taken the limit $D_F(2r)\rightarrow \delta^4(r)/16M_0^2$, and the field $\widetilde{\Phi}(r)$ appears here.
Then $\widehat{g}_Y\widetilde\Phi(0) = \widehat{g}_Y\phi_\omega(0)=g_Y$,
hence we revert to the standard model:
\bbo 
\backo
S_Y\rightarrow  {g}_Y M_0^2 
\int\!\! d^4X \;[\bar{\psi}_{iL}(X)\psi_{R}(X)]_{f}\;H^i(X) {+h.c.}.
\ebo
We see that $g_Y$ is nonrenormalized. 
Since the classical average field, $\widetilde{\Phi}(0)=\phi_\omega(0)$, appears here it imples that the result obtained for $M_0$ from the
solution to the SKG equation remains intact, i.e., $M_0\sim 6$ TeV with $|\mu| \sim 88$ GeV.
If we don't take the strict $\delta$-function limit we can do an expansion of 
the integrand in $r^\mu$.  This generates a series of higher dimension operators in
inverse powers of $M_0^2$
providing potentially sensitive probes to $M_0$ and the shape of the wave-function $\Phi(r^\mu)$,
briefly described below in section IV.

\vspace{0.1in}
\noindent  {\bf $S_\lambda$: }
The quartic term likewise becomes,
\bbo
\backo\backo  S_\lambda=
-\half\hat{\lambda}\int d^4X d^4r\; \delta^4(r) | H'(X)\Phi(r^\mu) |^4
=
-\half\hat{\lambda} \int d^4X  |H(X^\mu)|^4 |\widetilde{\Phi}(0)|^4 =-\half{\lambda} \int d^4X  |H(X^\mu)|^4, 
\ebo
where $\lambda  =\hat{\lambda}|\widetilde{\Phi}(0)|^4 $ is therefore not renormalized. Essentially $\lambda $ is determined by
the Yukawa coupling, $g^4_Y$, through the usual renormalization group running $\lambda$ from $M_0$ to $|\mu|$
and the successful result  obtained
from the underlying theory at one loop, e.g., $\lambda\approx 0.23$, is not modified.
This term would also permit an expansion in $r^\mu$ and yield an operator expansion of new physics.
 
 \vspace{0.1in}
 Finally, the result of inserting solutions to the SKG equation, integrating the
kinetic term by parts, yields the eigenvalue of the full action:  
\bbo
S_\Phi =-Z'M_0^4 \mu^2 \int d^4X d^4r |H'|^2 \;  |{\Phi}|^2 = -\mu^2\!\! \int d^4X |H|^2,
\ebo
where we have with $Z'H' = Z H$, and,  for any $\omega_i$:
\bbo
\label{normnn}
ZM_0^4\int \!\! d^4r |\Phi(r^\mu)|^2 =ZM_0^4\int \!\! d^4r |\phi_{\omega_i}(r^\mu)|^2 = 1,
\ebo
We thus obtain the same eigenvalue for the action with  $\Phi(r^\mu)$ as  for the action with  the underlying   $\phi_\omega(r)$.


It is straightforward to repeat the above with a continuous, manifestly Lorentz invariant, integral representation.
To match definitions used above we require the matching condition to the classical collective field, $\Phi(r^\mu)$, 
betweeb the discrete and continuous representations,
\bbo
\label{41}
\Phi(r^\mu) = \frac{1}{\sqrt{N}}\sum_i^N\phi_{\omega_i} =  {\cal{N}}\int d^4\omega\; \delta(\omega^2-1)\;\phi_\omega(r^\mu) 
\equiv {\cal{N}}\int_\omega \;\phi_\omega(r^\mu).
\ebo
To apply this, consider 
the expression  for the normalization of $\Phi$ above:
\bbo
1=
ZM_0^4\int  d^4r |\Phi(r^\mu)|^2 =
ZM_0^4\int d^4r\;{\cal{N}}\int_\omega \phi^\dagger_{\omega} (r)\;
{\cal{N}}\int_{\omega'} \phi_{\omega'}(r)
\nonumbo
=
ZM_0^4\int d^4r \; {\cal{N}}^2\int_\omega \phi^\dagger_{\omega} (r)
\phi^\dagger_{\omega}(r) =ZM_0^4\int d^4r \frac{1}{N}\sum \phi^\dagger_{\omega} (r)\phi_{\omega}(r),
\ebo
 where we use the orthogonality relation, eq.(\ref{240}), 
hence we have: $ 1={\cal{N}}^2N  $.

 Note that $\cal{N}$ is the normalization of  a divergent integral over the hyperboloid, and requires, in principle,
 regularization.  We will not enter into a detailed discussion of regularization here, 
however we note that we
can usually perform a ``Wick rotation.''
If the integral is an analytic function of the metric,
$g_{\mu\nu}= (1,-1,-1,-1)$, it can be continued as $g_{\mu\nu} \rightarrow \eta_{\mu\nu}\sim (1,1,1,1)$.
Then the hyperboloid is replaced by a Euclidean 4-sphere and
\bbo
{\cal{N}}\int \!\! d^4\hat{\omega}\;\delta(1-\hat\omega^2) =
{\cal{N}}\pi^2\int \hat\omega^2 d \hat\omega^2\delta(\hat\omega^2-1)
=\pi^2{\cal{N}},
\ebo
We then replace $\eta$ by $g$. Moreover, alternative averaging functions could be defined.
For example, if we define $\phi_\omega(r^\mu) =  \phi(\omega^\mu(\omega\cdot r) - r^\mu)$
then we could take $\int_\omega\rightarrow \int d^{4-\epsilon}\omega$ as in a momentum integral,
and thus use dimensional regularization.


\section{Full Action}

\subsection{Recovery of Standard Model in Pointlike Limit of the Potential}

The full action in the pointlike limit {\em of the potential}, $D_F (2r)\rightarrow \delta^4(r)/16M_0^2$, then becomes that of the standard model BEH boson coupled
to third generation (top)  quarks with the unprimed parameters.  While $\Phi(r^\mu)$ remains extended, it
is integrated out, by its normalization, and relative time does not appear:
\bbo
S =
\!\!\int\!\!d^4X \; \bl |DH(X^\mu) |^2 
-\mu^2 |H|^2     
 + g_YH^{i\dagger}(X)[\bar{\psi}_{R}(X)\psi_{iL}(X)+h.c.]_f -\half\lambda (H^\dagger H)^2 \br  +S',
 \ebo
 where $S'$ contains the free (unbound) top quark action and interactions through coloron exchange.
 
The theory generates the usual ``sombrero potential'':
 \bbo
\mu^2 |H|^2     +\half\lambda (H^\dagger H)^2,\qquad \text{where,} \;\;\; \mu^2 < 0.
 \ebo
 We extremalize the sombrero potential to obtain the broken phase, i.e., for the vacuum of the SM,
we therefore find:
\bbo
 H'(X^\mu){\Phi}(r) = H(X) \widetilde{\Phi}(r) \rightarrow \exp(i \pi^a(X)\tau^a/2v_{weak}) \bl
 \begin{array}{c} v_{weak} + \frac{h(X)}{\sqrt{2}} \\ 0 \end{array}\br \widetilde{\Phi}(r),
\ebo
where $h(X)$ is the physical BEH boson and $\pi^a(X) $ are Nambu-Goldstone bosons,
which are phase factors of $H(X^\mu)$.  The electroweak symmetry is broken spontaneously and
the gauge fields absorb the Nambu-Goldstone phase factors and acquire mass  in the usual way.  The resulting broken
phase of the BEH field, $h(X)$, is then as in the standard model
and the canonical normalization of $h(X)$ follows from the $Z$ normalization
of the internal field ${\Phi}(r)$ in eq.(\ref{S1}).
 Hence the SM BEH field, $h(X)$, is a collective object and its precise two-body nature is
 actually blended with the collective vacuum $\Phi$.  Likewise, the Nambu-Goldstone
 bosons are ``eaten'' by the gauge fields in the usual way.

The presence of the collective wave-function $\Phi(r^\mu)$ is not detectable in
the kinetic terms, nor in the  interaction in the pointlike limit of the potential.  
 We  obtain the BEH kinetic term from eq.(\ref{Zr}) as in the SM:
 \bbo
\label{HiggsK}
\backo\backo S_1\rightarrow 
\half ZM_0^4\!\!\int\!\!d^4X\;d^4r\;  (\partial h(X))^2 |{\Phi}(r)|^2  = \half \int\!\!d^4X(\partial h)^2,
\ebo 
where we integrate out ${\Phi}$ using eq.(\ref{normnn}). In fact, there is no way to discern the compositeness
of the BEH field from the kinetic and mass terms (this requires the Yukawa interaction and quartic terms below). 
Moreover, $\langle H \rangle \rightarrow v_{weak}$ with the covariant derivative, $D_\mu$, of eq.(\ref{covd}),
and eq.(\ref{normnn})
leads to:
\bbo
\label{Wr}
\backo\backo S_1=M_0^4\!\!\int\!\!d^4X\;d^4r \bl Z|D H(X^\mu) |^2|{\Phi}(r) |^2\br \rightarrow
\nonumbo
= ZM_0^4\!\!\int\!\!d^4X\;d^4r  \bl M_W^2 W^+ W^- + \half M_Z Z^2\br|{\Phi}(r)|^2  = \!\int\!\!d^4X
\bl M_W^2 W^+ W^- + \half M_Z Z^2\br.
\ebo
Hence we generate the $W^\pm$ and $Z^0$ mass terms in the usual way, where the Nambu-Goldstone bosons
have become their longitudinal gauge components.
The usual effective action for the BEH boson $h(X)$ emerges:
\bbo
\int d^4 X\bl
\half \partial_\mu h \partial^\mu h -\half m_h^2 h^2 - \sqrt{\frac{\lambda}{2}}|\mu|  h^3 - \frac{\lambda}{8}h^4 \br+ \text{electroweak couplings},
\ebo
where $m_h=\sqrt{2}|\mu|= 125$ GeV.

\subsection{Integrating Out $r^\mu$ and Higher Dimension Operators}

The structure of the collective field, $\Phi(r^\mu)$, can in principle be probed through the extended Yukawa and
quartic interactions,
but the effects will be suppressed by powers of $M_0$. 
The Yukawa coupling term in natural top condensation takes the form \cite{chris1,chris2} in the broken
phase of eq.(\ref{yukyuk}) (recall there is no $Z'$ factor on the interactions and $H'\Phi = H\widetilde{\Phi}$):
\bbo 
\label{Yuk2}
\backo
S_Y'
=  \widehat{g}_Y M_0^2
\int\!\! d^4X d^4r \;[\bar{\psi}_L(X\!+\!r)\psi_{R}(X\!-\!r)]_{f}D_F(2r^\mu)\; \bl v_{weak} + \frac{h(X)}{\sqrt{2}}\br\widetilde{\Phi}(r^\mu) {+h.c.},
\ebo
In the pointlike approximation  of the potential 
and we have  $ D_F(2r^\mu)\rightarrow (16M_0^2)^{-1} \delta^4(r)$, then $\widehat{g}_Y\widetilde\Phi(0) = \widehat{g}_Y\phi_\omega(0)=g_Y$, which then leads to the conventional mass term for the top quark and BEH coupling:
\bbo 
\label{Yuk3}
\backo
S_Y'
=  g_Y  
\int\!\! d^4X  \;[\bar{\psi}_L(X)\psi_{R}(X]_{f}\bl v_{weak} + \frac{h(X)}{\sqrt{2}}\br {+h.c.}
\qquad \text{where,}\qquad m_{top} = g_Y v_{weak}.
\ebo

If we go beyond the pointlike limit we can integrate out $r^\mu$ and
the BEH-Yukawa interaction will 
generate  $O(1/M_0^{2p})$ corrections.  Note that if only a single component field, $\phi_\omega (r)$,
comprised the entire wave function, e.g., $\Phi(r)=\phi_\omega(r)$, then we would 
generate Lorentz violating higher dimension operators at this stage.
However, with the collective field $\Phi(r)$ that sums over all $\phi_{\omega_i}(r)$,
as in eq.(\ref{defdef}), these effects are now
{\em bona fide} Lorentz invariant.

Rather than taking the pointlike potential limit,  
we can rewrite eq.\ref{Yuk}) in the broken phase as:
\bbo 
\label{Yuk}
\backo
S_Y' =m_{top}
\int\!\! d^4X d^4r \;[\bar{\psi}_L(X\!+\!r)\psi_{R}(X\!-\!r)]_{f}(M_0^2D_F(2r^\mu))\;G(r^\mu)+h.c.,
\qquad G(r^\mu)= \frac{\widetilde{\Phi}(r)}{\widetilde{\Phi}(0)}\sim 1+ a_1 r^2 + ...
\ebo
Likewise, we can expand $[\bar{\psi}_L(X\!+\!r)\psi_{R}(X\!-\!r)]\sim \sum_{p}c_{p}r^p\bar{\psi}_L(X) D_X^{p}\psi_{R}(X) $
where $D_X$ are covariant derivatives wrt $X$.
Hence the integrand becomes a power series in $r^p$. Then, integrating by parts, we have: 
\bbo 
\label{}
\backo
\int d^4r \;
C_p(X) r^p D_F(2r^\mu)\sim \int d^4r\int\!\! \frac{d^4q}{(2\pi)^4}\;C_p(X) e^{iqr} \frac{\partial^p}{\partial q^p}\bl\frac{1}{q^2-M_0^2}\br \sim 
C_p(X)
\frac{1}{M_0^{2+p}},
\ebo
where we take the pointlike limit in the last term. Hence,
 the expansion in $r$ corresponds to an expansion in $M_0^{-1}$.
The expansion in $r^\mu$ in all terms in the integrand generates a tower of assorted Lorentz invariant 
higher dimension operators
such as a leading term: 
\bbo
\frac{m_t}{M_0^2}\bl 1 + \frac{h(X)}{\sqrt{2}v}\br\bl[\bar{b}_L(X)D^2t_{R}(X)] + [\bar{t}_L(X)D^2t_{R}(X)]+...\br+h.c.,
\ebo
where $D^2$ is the covariant derivative, including  the gluon, $\gamma$, $W^\pm$,  $Z$ 
couplings.
These operators represent new contact terms and processes such as:
\bbo
\bar{t}\rightarrow b +W + (g,\gamma,Z)\qquad \text{and,} \qquad 
\bar{t}\rightarrow t  + (g,\gamma,Z).
\ebo
These can be probed in decays or in
production in, e.g., a lepton collider via: 
\bbo
(\ell^+\ell^-) \rightarrow (\gamma^*, Z^*) \rightarrow t+b+(W,g,\gamma,Z).
\ebo
Determining the full set of effective operators is straightforward, but beyond
the scope of the present paper.


\section{Summary }

We have proposed a nontrivial vacuum for the natural top condensation theory.
The vacuum is manifestly Lorentz invariant, composed of a collective Lorentz invariant
sum over internal wave-functions, $\Phi = (1/\sqrt{N})\sum_\omega \phi_{\omega}(r^\mu)$. 
Though each internal wave-function, $\phi_{\omega} $,
is independent of relative time, $\tau$, they each have  dependence upon the arrow
of relative time, $\omega^\mu$, and are  solutions to the
Schr\"odinger-Klein-Gordon equation with eigenvalue $-|\mu^2|$. 
The sum over $\phi_{\omega} $ makes the collective field, $\Phi(r^\mu)$, Lorentz invariant and 
independent of  $\omega^\mu$.  Residual dependence upon $r^\mu$ can then be integrated out leading to
suppressed Lorentz invariant operator corrections.

The bilocal BEH field in the vacuum is then a minimum of the sombrero potential
and takes the form:
\bbo
H'(X)\Phi(r^\mu)
= H(X^\mu)\widetilde{\Phi}(r)  \rightarrow \exp(i \pi^a(X)\tau^a/2v_{weak}) \bl\begin{array}{c} v_{weak}\widetilde{\Phi}(r^\mu) + h(X,r^\mu) \\ 0 \end{array}\br,
\ebo
where $v_{weak}= |\mu|/\sqrt{2}$ and the observed BEH  boson is $h(X)$ where:
\bbo
h(X,r^\mu) =h(X)\widetilde{\Phi}(r^\mu).
\ebo

 Much can be done to explore this theory further, including the bilocal formalism itself.
 For a more complete theory, we note that
 the $b_R$ quark also participates in topcolor in an anomaly-free scheme, {\em e.g.}, see \cite{Topcolor},
 and this should be adopted into the present dynamics.
 An additional $Z^{0}{'}$ interaction can be introduced that makes the $\bar{b}b$ 
 channel sub-critical, hence non-binding.

 The theory is therefore  testable, 
 mainly by direct discovery of the octet of colorons at $M_0\approx 6$ TeV.
 The theory points toward an $SU(3)'\times SU(3)\times SU(2)\times U(1)_Y$   gauge structure emerging at the $\sim 10$ TeV scale with likely additional $U(1)'s$.
 The theory also offers sensitive probes of new contact interactions involving $t$- and $b$-quarks, and
 though flavor mixing there may be induced rare processes involving the other quarks and leptons \cite{Technicolor}.
 
 The $\bar{b}_Lb_R$ mass term can in principle arise from instantons in the $SU(3)'$ sector
 as in \cite{Topcolor}; the third generation lepton $(\tau)$ mass may arise by extending $SU(3)'\rightarrow SU(4)$,
 i.e., treating leptons as the fourth color.
 Light fermion masses are presumably then generated in analogy to extended technicolor models \cite{Technicolor}.
 So far, we have relied upon the intuition
 of the 1990's topcolor scheme, \cite{Topcolor}, but a more complete natural top condensation theory, including the light particle masses can be formulated.  Our theory implies, therefore, a novel gauge dynamics, compositeness,  and possibly the entry point
 to a complete theory of flavor physics.

Note that bilocality of the wave-function is an important naturalness constraint. 
One might be tempted to, e.g., 
``loop'' the Yukawa  interaction and argue for a problematic large correction to the BEH mass $\propto -\bar{g}_Y^2 N_c M_0^2$.
This would lead to the {\em false conclusion} that the effective theory is  ``unnatural.''
The loop actually generates an enhancing correction to the bilocal potential,
i.e., $\sim g_Y^2 \delta^4(2r^\mu)/M_0^2 \sim g_Y^2 D_F(2r^\mu) $ for large $M_0$ \cite{chris2},
that leads to a ``critical amplification'' of the
 effective  4-fermion coupling,
$\bar{g}^2_0$, where:
$
\bar{g}_0^2 = g_0^2\left(1-{g_0^2N_c}/{8\pi^2} \right)^{-1}.
$
Hence, while $\bar{g}_0^2$ is  supercritical, 
the underlying topcolor coupling, $g_0^2$, is smaller and subcritical. 
 
In conclusion, we emphasize this theory is natural and manifestly Lorentz invariant.
The approximate scale symmetry near
critical coupling 
provides the custodial symmetry of the small $|\mu|^2$. 
The low energy physics is controlled by the $\Phi(r^\mu) \sim \phi(r^\mu)$ 
wave-function spreading,
rather than the renormalization group of \cite{BHL}.
The hierarchy is protected against additive radiative corrections
by the bilocality, i.e., there is no ``additive quadratic divergence,'' 
but only additive and enhancing renormalizations of the 
bilocal binding interaction \cite{chris2}.  We used
the source/Legendre-transform methods of Jackiw et.al., \cite{Jackiw}, 
to derive the effective semiclassical theory used here, which leads to critical amplification of the potential
coupling $\bar{g}_0^2$ \cite{chris2}. 
The fine-tuning is at the few $\%$ level, and, indeed, this may be the first and only minimally-fine-tuned 
theory
of the BEH boson that is consistent with experiment and testable in the not-too distant future.

\appendix

\section{Orthogonality of $\phi_{\omega_i}(r)$}

Recall that
the $\phi_{\omega_i}(r^\mu)$ solutions of the Lorentz invariant 
integro-differential equation are normalized as in eq.(\ref{newnorm}):
\bbo
1=ZM_0^4\!\! \int\! d^4r \; \phi_{\omega_i}^\dagger(r) \phi_{\omega_{i}}(r) =
1=M_0^3\!\! \int\! (ZM_0)\omega_{\mu i} dr^\mu\; d^3r_\perp  \; \phi_{\omega_i}^\dagger(r) \phi_{\omega_{i}}(r).
\ebo
The latter Lorentz invariant expression can be evaluated in the rest frame, where $\omega_\mu = (1,0,0,0)$:
\bbo
\label{24}
1= M_0^3\!\! \int\! d^3r |\phi_{\omega}(\vec{r})|^2, \qquad\text{where,}\qquad 1=ZM_0\int dr^0 \equiv ZM_0T.
\ebo
In the small $|\mu|<\!\!M_0$ limit the normalization integral eq.(\ref{newnorm}) is dominated by large $r$, 
and we have the large distance solution in the rest frame, $\phi(r)\sim Ne^{-|\mu|r}/r$, hence:
\bbo
1= M_0^3\!\! \int\! 4\pi r^2 dr \frac{N^2e^{-2|\mu|r}}{r^2} \sim 2\pi \frac{N^2 M_0^3}{\mu},\qquad \text{hence,}
\!\!\qquad\!\! N^2= \mu/2\pi M_0^3.
\ebo
If we consider $Z\sim 1/M_0T<\!\!<1$, then $M_0\int \omega_\mu dr^\mu \sim  M_0 T >\!\!>1$, then the $ \phi_\omega^\dagger(r^\mu)$
become orthogonal in $\omega$,
\bbo
\int d^4 r\; \phi^\dagger_\omega(r)\phi_{\omega'}(r) = 0, \qquad \omega_\mu \neq \omega'_\mu.
\ebo
To see  orthogonality, consider the timelike hyperboloid defined by $\omega^2=1$ and 
choose $ \omega_0=(1,0,0,0)$,  and $ \omega'=(\cosh\theta,\sinh\theta, 0,0)$
where $\theta$ defines a boost in the $x$ direction.
Then $\phi_{\omega_0}(r^\mu)\rightarrow \phi(0,r_x,r_y,r_z)$, with $r^0$ as 
the flat direction for $ \omega_0$, and $\vec{r}=(r_x,r_y,r_z)$.  
Then,
\bbo
\phi_{\omega'}(r^\mu)= \phi(\omega'^{\mu}\omega'_\nu r^\nu-r^\mu  ) =\phi(r^0\sinh^2\theta, r_x\cosh^2\theta, r_y,r_z ).
\ebo
We consider small $\theta$, hence  in leading order in $ \theta^2$ it is sufficient 
to use $\phi_{\omega'}(r^\mu)\approx \phi(r^0\theta^2 , \vec{r} )$.
The overlap integral is dominated by the large $r=|\vec{r}|$ component and in this limit with eq.(\ref{24}), 
and flat direction $r^0$, using a cut-off $r^0 =T$ on the integral:
\bbo
1=ZM_0^4\!\! \int\! d^4r \; \phi_{\omega_0}^\dagger(r) \phi_{\omega'}(r)
=ZM_0^4\!\! \int\! 2\pi  r^2 dr dr^0   
\frac{N^2e^{-2|\mu| r}}{\sqrt{{r}^2((r^0)^2\sinh^2\theta + {r}^2 )}} 
\nonumbo
\approx \frac{\pi Z{N^2}M^4_0}{\theta \mu^2} \ln(M_0T ) = \frac{1}{2\mu\theta T} \ln(M_0T ).
\ebo  
In the $T\rightarrow \infty$ limit this approaches zero.

The result is not identically zero.  In the
vacuum, however, where these fields will be clustered into a stable collective state
and the system cannot decay, then $T$ can go to infinity with impunity.

\section{Brief Summary of Natural Top Condensation  }

Our formalism, ``Natural Top Condensation,'' \cite{chris1,chris2}, is Lorentz invariant and postulates an attractive 
``topcolor'' interaction \cite{Topcolor} of strength $g_0^2$ at 
a high scale $M_0$. The bound states are correlated pairs $\bar{\psi}(y)_L\psi_R(x) \rightarrow
\Phi(x,y)\sim \Phi(X,r)$, and we follow Yukawa in writing invariant
kinetic terms for the pairs, consistent with their single particle kinetic terms \cite{Yukawa}.
This  
yields a  bound state given by a Schr\"odinger-Klein-Gordon (SKG) equation
satisfied by an internal wave-function, $\phi(r)$. This has eigenvalue $\mu^2$, which is the Lagrangian mass
of the BEH boson and we have  a Yukawa interaction between the bound state and unbound fermions. 
For supercritical coupling, $g_0 > g_{c}$ we find $\mu^2 < 0$, and which implies spontaneous symmetry breaking.
For small $|\mu|<M_0$, near critical coupling, we have significant wave-function spreading 
and ``dilution'' of $\phi(0)\sim \sqrt{|\mu|/M_0}$. The
resulting top quark Yukawa, $g_Y\propto \phi(0)$, and quartic couplings,
$\lambda\propto |\phi(0)|^4$, are subject to
power law suppression, rather than the relatively slow renormalization group (RG) evolution
in the old Nambu--Jona-Lasinio (NJL \cite{NJL}) based top condensation model. 
The dilution effect significantly reduces the hierarchy and,
remarkably, the standard model (SM) quartic coupling, $\lambda\approx 0.25$, 
becomes concordant with experiment.   The 
fine tuning of the model is also vastly reduced by dilution to  $\sim \phi(0)^2 \sim |\mu|/M_0 \sim$ few $ \%$.
Our central prediction is the existence of a binding interaction due to a color
octet of massive gluon-like objects, called ``colorons,''  \cite{Topcolor}\cite{NSD}\cite{colorons}, with mass $M_0\sim 6$ TeV.
The colorons may be accessible to the LHC. Moreover, loop effects enhance the binding in the $0^+$ channel, and the
requisite $\mu^2<0$ can occur for significantly weaker coloron coupling \cite{chris2}.
The top condensation model of a composite BEH boson therefore becomes a
compelling theory. Inputting the induced Yukawa coupling $g_Y\approx 1$ we obtain
the resulting prediction $M_0\sim 6$ TeV.
This construction was confirmed by applying the formal
source/Legendre-transformation methods of Jackiw, {\em et. al,} \cite{chris2}\cite{Jackiw}.

Starting with  third generation fermions, $\psi_{L,R}$,
coupled to a  coloron exchange potential, we obtain
an effective, Lorentz invariant interaction structure for the bilocal BEH boson $H(X^\mu)\phi(r^\mu)$ in the symmetric phase
of the standard model (SM) \cite{chris1}. This was independently
derived using techniques of Jackiw, and Cornwall, Jackiw and Tomboulis,
\cite{chris2}\cite{Jackiw}.
The kinetic terms follow by electroweak gauge invariance, and 
in ref.\cite{chris1} we introduced  Wilson lines to ``pull-back'' the electroweak 
gauging to $X$. Hence the covariant derivative of $H$ becomes the standard BEH form:
 \bbo
 \label{covd}
 D_{H\mu}=\frac{\partial}{\partial X^\mu}- ig_2W^A(X)_\mu \frac{\tau^A}{2}-ig_1 B(X)_\mu \frac{Y_H}{2}.
 \ebo
With the pullback, $\phi(r)$ is a dimensionless complex scalar and has no
gauge charges. The Wilson line pullback  is essentially
 a low energy approximation for the electroweak interactions, but makes the effects of symmetry breaking
transparent.
-
In the barycentric coordinates we have:
\bbo
\label{Haction10}
\backo\backo
S=
 M_0^4\!\!\int\!\!d^4X\;d^4r \bl Z|D_{H}H(X^\mu) |^2|\phi(r) |^2
+Z|H(X^\mu)|^2|\partial_{r}\phi(r) |^2 +2{g_0^2 N_c}D_F(2r)|H^\dagger H||\phi(r)) |^2\br \!\! +S_Y+S_\lambda + ...
 \ebo   
where the $H(X^\mu)$ kinetic term is canonical with 
the Lorentz invariant normalization of $\phi(r)$ is:
\bbo
\label{Lnorm}
1 =
M_0^4 Z\int \!\! d^4r |\phi(r^\mu)|^2\rightarrow M_0^3 \int \!\! d^3r |\phi(r^\mu)|^2,
\ebo
where  $Z$ is as defined upon removal of relative time as in (\ref{line}). The Yukawa interaction is also generated at tree level,
\bbo 
\label{SY}
\backo
S_Y 
= \widehat{g}_Y M_0^2
\int\!\! d^4X d^4r \;[\bar{\psi}_{iL}(X\!+\!r)\psi_{R}(X\!-\!r)]_{f}D_F(2r)\;H^i(X)\phi(\vvr) {+h.c.},
\ebo
and a quartic interaction is generated at loop level, given in the point-like $\phi(r)$ approximation by:
\bbo 
\label{Slam}
\backo
S_\lambda 
=  -\frac{\widehat{\lambda}}{2}
\int\!\! d^4X  (H^\dagger H)^2 |\phi(0)|^4 {+h.c.}. = -\frac{\lambda}{2}\int\!\! d^4X  (H^\dagger H)^2  {+h.c.}.
\ebo
In the above, the Yukawa and quartic couplings  $g_Y$ and $\lambda$ are derived quantities 
from the underlying theory. In a pointlike approximation for the interactions, $D(2r)\sim \delta^4(r)/M_0^2$,  we have.
\bbo
\label{A62}
g_Y\approx  g_0^2 \sqrt{2N_c/J}\;\phi(0),
\qquad\qquad
\lambda \approx (g_Y^{4}-g_Y^2\lambda) \frac{N_{c}}{4\pi ^{2}}%
 \ln \left( \frac{M_0}{\mu }\right) .
\ebo
Note that $g_Y$ is classical and  $\lambda$ arises at loop level (${\cal{O}}(\hbar)$). 
We thus see that these are subject to dilution in an extended solution with the internal wave-function  $\phi(r)$
and $g_Y\propto \phi(0)$ and $\lambda\propto | \phi(0)|^4$.
Experimentally we have $g_Y \approx 1$ and $\lambda \approx 0.25$.

The full action then  takes the form:
\bbo
\label{Haction2}
S =
\!\!\int\!\!d^4X \; \bl |D_{H}H(X^\mu) |^2 
+|H(X^\mu)|^2 S_\phi+    
 \ g_YH^{i\dagger}(X)[\bar{\psi}_{R}(X)\psi_{iL}(X)+h.c.]_f -\half\lambda (H^\dagger H)^2 \br  +S',
 \ebo
 where $[...]$ denotes color indices are contracted, and $i$ is an $SU(2)_{weak}$ index.
Here $S_\phi$ describes the internal wave-function field $\phi(r^\mu)$, and $S'$ describes
the coupling of the bound state to external free fermions:
 \bea 
 \label{bcs}
\backo 
S_\phi & = & M_0^4\int \!\! d^4r \bl Z \partial_\mu\phi^\dagger(r) \partial^\mu \phi(r) + 2g_0^2N_c D_F(2r^\mu) \phi^\dagger(r)\phi(r)\br
\\ 
 \backo 
S' &=&
\!\int\! d^4x \bl  
 [\bar{\psi}_L(x)i\slash{D}\psi_{L}(x)]_f+ [\bar{\psi}_R(x)\;i\slash{D}\psi_{R}(x)]_f\br +g_0^2\!\int\!d^4x d^4y \; [\bar{\psi}^i_L(x)\psi_{R}(y)]_fD_F(x-y)[\bar{\psi}_{R}(y)\psi_{iL}(x)]_f .
\eea
In $S'$ we have free unbound fermions, $\psi_{Lif}\sim (t,b)_L$ and $ \psi_{Rf}\sim t_R$
(the minimal model omits $b_r$ but this can be readily incorporated as in \cite{Topcolor}).
Note that the internal field action $S_\phi$ is nested within the
action for a conventional pointlike BEH boson, $H(X^\mu)$.
$D_F(2r^\mu)$ is the Feynman propagator function for the massive colorons:
\bbo 
\label{DF}
D_F(x-y) =- \int\frac{1}{q^2-M_0^2}e^{2q_\mu r^\mu}\frac{d^4q}{(2\pi)^4},
\ebo 
where $r^\mu $ is a radius hence the factor of $2r^\mu$.
Note the BCS-like enhancement factor of $N_c$ in eq.(\ref{bcs}))
in the $\phi^\dagger D_F(2r^\mu) \phi$ interaction term. 
 
The value of $M_0$ is then determined by inputting $g_Y = 1 $, 
 and the  known value of the symmetric phase (Lagrangian mass) of the BEH boson,
 which is 
  $-|\mu|^2 =-(88)^2$ GeV$^2$.  We find that {\em the scale $M_0$ is predicted,
  $M_0\approx 6$ TeV } and is no longer the nonsensical $10^{15}$ GeV in the old top condensation
  based upon the NJL-model \cite{BHL}.
  This is due to the faster power-law 
  running of $g_Y \propto \phi(0)\sim \sqrt{|\mu|/M_0}$, rather than the slow, logarithmic 
  RG running of $g_Y$ in the NJL model.
  
  Moreover, a stunning result of this model is the quartic coupling $\lambda$.
Experimentally, in the SM using the value of $m_{BEH}\approx 125$ GeV and $v_{weak}\approx 175$ GeV 
we find $\lambda \approx 0.25$.
In the present bilocal scheme, owing to dilution of $\phi(0)$, the quartic coupling
is also suppressed  and is now generated in RG running  from a value
of $\lambda(M_0) =0$ at $M_0 \approx 6$ TeV, down  to $\lambda(|\mu|)$ with $|\mu|\sim 88$ GeV, using
$g_Y\approx 1$.  
The prefactor at one loop level reflects the full RG running of $\lambda$,
and at leading log the RG equation for $\lambda$ yields \cite{RGlam}:
\bbo
\label{A11}
\lambda \approx (g_Y^{4}-g_Y^2\lambda-[\lambda^2]) \frac{N_{c}}{4\pi ^{2}}%
 \ln \left( \frac{M_0}{\mu }\right) \approx 0.23\;\;\; \makebox{(cf., $0.25 $ experiment.)},
\ebo  
where we solve for $\lambda$ self-consistently with $g_Y=1$ and $M_0\sim 6$ TeV.
Note that the $[\lambda^2]$ term should be omitted since it involves internal
propagation in loops of the composite BEH boson (and only slightly affects the result). 
The $g_Y^2\lambda$ terms are fermion loop leg renormalizations.
This is  in  excellent agreement with experiment at one loop precision
and significantly contrasts the prediction of the old NJL-based top condensation model where the quartic coupling
was determined by the RG and found to be $\lambda\sim 1$ \cite{BHL}, much too large.
  
 The degree of fine-tuning of the theory is remarkably suppressed by $|\phi(0)|^2$ in
  a subtle way. Rather than the naive result one would expect from the NJL model, 
  $\delta g_0^2/g_c^2 \sim|\mu|^2/M_0^2 \sim 10^{-4}$,
  we now obtain a linear relation: $\delta g_0^2/g_c^2 \sim |\phi(0)|^2 \sim |\mu|/M_0 \sim 1\%$.
  This is derived in \cite{chris1}, but was accidentally noticed when
  the bound state was treated by a variational ``spline approximation'' in
  earlier papers \cite{chris2}.

 \section*{Acknowledgments}
 
  I thank Bill Bardeen, Bogdan Dobrescu, Tao Han and other participants of Photon-Lepton 2025,
  for discussions and comments.   Sadly, Bill Bardeen passed away during the completion of
  this work. His awesome intellect, depth of understanding of physics, and close friendship, are sorely missed.


\end{document}